# A repulsive skyrmion chain as guiding track for a race track memory

D. Suess[1*], C. Vogler[2], F. Bruckner[1], C. Abert[1]

[1]Doppler Laboratory "Advanced Magnetic Sensing and Materials", University of Vienna, Währinger Straße 17, 1090, Vienna.

[2]Physics of Functional Materials, University of Vienna, Währinger Straße 17, 1090, Vienna.

***Abstract:*** A skyrmion racetrack design is proposed that allows for thermally stable skyrmions to code information and dynamical pinning sites that move with the applied current. This concept solves the problem of intrinsic distributions of pinning times and pinning currents of skyrmions at static geometrical or magnetic pinning sites. The dynamical pinning sites are realized by a skyrmion carrying wire, where the skyrmion repulsion is used in order to keep the skyrmions at equal distances. The information is coded by an additional layer where the presence and absence of a skyrmion is used to code the information. The lowest energy barrier for a data loss is calculated to be ∆E = 55 $k_B T_{300}$ which is sufficient for long time thermal stability.

---

[1]* Correspondence to dieter.suess@univie.ac.at

Skyrmion racetrack devices that store information by the presence or absence of skyrmions have recently attracted significant attention [1–4]. Extensive research has been done in order to calculate the stability of individual skyrmions for infinitely extended magnetic structures. Using variants of the nudged elastic band method [5], that is well established for magnetic structures [6], the energy barrier for skyrmion annihilation via Bloch points was studied recently [7,8]. There is scientific debate whether these structures are topologically protected in infinite systems in the continuum limit. If they are protected, the change of the topological structure of the skyrmion with topological charge one to the homogenous state with topological charge zero is not possible without surpassing an infinite energy barrier [9]. In Ref[10,11] it is argued that this transformation is possible in continuous systems via finite barriers . However, for practical application the path of annihilation via the boundary is more relevant because this path usually requires significant smaller energy barriers to annihilate the skyrmion [3,12,13].

The stability of Skyrmions against annihilation is often considered to be the most important feature for storage applications. However, this kind of stability is not a sufficient condition for stable data. Since in skyrmion racetrack memory the information is coded by the position of the skyrmion it has to be guaranteed that (i) the position of the skyrmion is thermally stable and (ii) if a current pulse is applied in a skyrmion racetrack shift register the distance between all the skyrmions remain the same.  Whereasrequirement (i) can be realized by pinning sites that can be constructed by geometrical or magnetic features [12],requirement (ii) is not solved with the original racetrack concept. Finite temperature and distributions in the pinning sites lead to a distribution in depinning currents and times that are expected to lead to read and write error rates that are significantly too large for practical applications [12]. If for example a too short current pulse is applied, the skyrmion may not move over the pinning site. If a too long current pulse is applied, the skyrmion may move over two pinning sites. For comparison, the required write error rates in random access memory must be smaller than $10^{-9}$ as reviewed in Ref[14]. In hard disk storage the write error rates are relaxed to be smaller than $10^{-2}$ due to extensive use of error correction codes that reduce the overall unrecoverable bit error rate to $10^{-16}$ - $10^{-14}$ as reviewed in Ref[15]. For comparison the switching time distribution divided by the mean switching time in MRAM cells is measured to be around 30% [16]. In order to achieve BER < $10^{-2}$ in skyrmion devices the switching time distribution divided by the mean switching time  must decreased to be smaller than about 5% [12].

In this letter we present a concept that solves the problem of intrinsic depinning time distribution of skyrmions. The concept utilizes two magnetic layers, which are separated by a non- magnetic layer as shown in Fig.  *1*. The bottom layer does not store information but it is used to realize moveable pinning sites. The repulsion of skyrmions [17] is used in order to keep the distance of these skyrmions constant. The information is coded in the top layer by the presence and absence of skyrmions. In order to move the information, a current is applied to the layers which moves the skyrmions. Alternatively the skyrmions can also be moved by spin orbit torque in an adjacent heavy metal layer where a current is applied. Due to the spin-Hall effect the spin polarization can be used in order to move the skyrmions in the top layers. Different designs are possible such as shifting the skyrmions in a ring structure. Due to the stayfield interaction the skyrmions in the top layer are always kept above the carrying skyrmions in the bottom layer as it will be shown in the following.  If the current is applied, the presence of the skyrmions in the guiding layer can be used as a clock cycle, i.e. whenever a skyrmion is detected in the guiding layer, the magnetic state in the information layer is determined in order to resolve the bit state. The problem, that the distance between the skyrmions may change in standard racetrack devices can be overcome with this device. Even if the distance between two skyrmions in one of the layers is changed due to defects, the guiding layer will eventually restore the original distances, which in turn readjusts the skyrmion distance in the information layer. This is in contrast to the original design of racetrack structures where an unintended change of distance can not be restored again. Compared to the interesting concept of Ref[18] it is expected that due to the planar magnetic structure the suggested device is significantly easier to fabricate. Furthermore, the data density is expected to be larger by a factor of two.

In the following, we calculate the thermal stability of the stored information by calculating energy barriers between the encoded bit state and various paths towards bit annihilation. For the micromagnetic simulations, a hybrid FEM / BEM method is used [19]. If not stated otherwise, the used mesh size is 3 nm. For the calculation of the energy barrier the string method is used [20]. In order to redistribute the images along the string we use for the reaction coordinate $s_i$ for the image $i$,

$$s_0 = 0, \quad s_i = s_{i-1} + W_{i-\frac{1}{2}} |\mathbf{u}_i - \mathbf{u}_{i-1}| \tag{1}$$

Here, $\mathbf{u}_i$ contains the 3$N$ Cartesian coordinates of the normalized magnetization, the used norm is the $L_2$ norm and

$$W_{i-\frac{1}{2}} = \left[ \left( \frac{E_i + E_{i-1}}{2} - E_{min} \right) / (E_{max} - E_{min}) + 1 \right]^{\omega}, \tag{2}$$

Where $E_i$ is the total energy of image i and $E_{max}$ and $E_{min}$ the maximum and minimum energy of the list of images, respectively. After 5ps of integrating the Landau-Lifshitz equation without precession term, the images are redistributed by a quadratic spline interpolation, in order to obtain equal distance according the reaction coordinate defined in Eq. (1). If not stated explicitly $\omega$ = 0, which simplifies the distance between the images to the $L_2$ distance.

The magnetic material parameters of the wire are: anisotropy constant $K_1$ = 0.6 MJ/m³, with the easy axis perpendicular to the film (z-axis), saturation polarization $J_s$ = 0.72 T, exchange constant A = 15 pJ/m. The DMI effect is included by a surface DMI field of the form,

$$\mathbf{H}_{DMI} = -\frac{\delta E_{DMI}}{\delta \mathbf{J}} = -\frac{2D}{J_s} \left[ \mathbf{n}(\nabla \cdot \mathbf{m}) - \nabla(\mathbf{n} \cdot \mathbf{m}) \right] \tag{3}$$

Here, **n** is the normal vector of the interface between the magnetic and heavy metal multilayers. The DMI constant of $D$ = 3 x 10$^{-3}$ J/m² is assumed to be independent from the repetition of the multilayer structure.. The magnetic parameters are effective parameters, where the values are averaged over the entire multilayer thickness.

In Fig. 2 the energy barrier for a path is calculated, where the skyrmion moves from one pinning site to the next one, as shown by the magnetic states in Fig. 2 (A) and Fig. 2 (D). Here, the z-component of the magnetization is color coded. Red represents magnetization up, blue represents magnetization down. The thickness of the guiding layer and the information layer is $t_{bottom}$ = 10 nm and $t_{top}$ = 5 nm, respectively. These two layers are separated by a 1 nm non-magnetic layer. The lateral dimensions are 180 nm x 90 nm. The energy barrier obtained by the string method is ΔE = 55 $k_BT_{300}$. The saddle point configuration is shown by Fig. 2 (B), where the skyrmion in the information layer partly overlaps with the second skyrmion in the gap protection layer. Interestingly during the motion of the skyrmion of the stable configuration Fig. 2 (A) to Fig. 2 (D) there exists also a local minimum as shown in Fig. 2 (C), where the skyrmion in the top layer increases in size and is pinned by the stayfield of the two skyrmions in the bottom layers. The energy barrier to leave the stable states Fig. 2 (A) and Fig. 2 (D) is sufficiently high to guarantee thermal stability at room temperature.

Another path that leads to data loss is shown in Fig. 3. Here, the energy barrier of annihilation of the skyrmion in the information layer is calculated. As initial states for the string method two states are chosen. The first state contains the skyrmion in the information layer and the two skyrmions in the gap protection layer (Fig. 3. (A)). The final state contains the same skyrmions in the gap protection layer but no skyrmion in the information layer (Fig. 3. (E)). Interestingly the skyrmion is transformed to an elliptical structure, where on the long axis of this structure the rotation of the spin direction is reversed (Fig. 3. (B) - Fig. 3. (C) ). The state with maximum energy along the MEP is shown in Fig. 3. (D), where the elliptical structure still has finite size. Finally A Bloch point is formed that has however for the used mesh size smaller energy than the state shown in Fig. 3. (D). Due to the formation of Bloch point, the micromagnetic approach is not valid any longer and it is expected that the results depend on the mesh size.

In order to study the mesh-size dependence of the annihilation process the model is simplified to only one magnetic layer with the dimensions of 90 nm x 90 nm x 0.6 nm, as shown in Fig. 4. For this simplified model without the gap protection layer the MEP is calculated. Fig. 4. shows the magnetic states for a mesh size of 1.6 nm. With increasing reaction coordinate along the minimum energy path the skyrmion diameter decreases in size as shown in Fig. 4 (A-B) until at the saddle point configuration a Bloch point is formed in Fig. 4 (C). A detailed view of the magnetization state is shown in Fig. 4 (C – detailed view). Here, the finite element mesh is shown, where the magnetization is represented by vectors on the node points At the saddle point four spins are pointing towards each other significantly contributing to the saddle point energy. This configuration which is obtained with the finite element method using a regular mesh is very similar to the state that is found by an atomistic simulation [7]. Simulations for a system with a smaller skyrmion diameter that can be tackled with atomistic models showed that the energy barrier difference of the skyrmion calculated with the micromagnetic model using an atomistic discretization compared to an atomistic simulation is only about 25%.

The saddle point energy and the energy barrier depends on the mesh size, since the exchange energy of the finite element that contains the Bloch point is determined by the discretization size. Fig. 5 (red circles) shows the energy barrier for annihilation via the described process as function of mesh size. An almost linear increase of energy barrier with the mesh size can be observed. If the mesh size is chosen similar to the atomistic lattice constant of $a$ = 0.5 nm the energy of annihilation via the Bloch point is about a factor 3.2 higher than the energy barrier via annihilation via the boundary. Hence, even if the calculation of annihilation via the Bloch point requires atomistic simulation, the finite element simulation using a discretization close to the atomistic mesh size gives an estimate. Since the barrier for annihilation via the boundary is significantly lower this annihilation path is expected to be the relevant one for the device. This barrier can be well resolved with micromagnetic simulations.

Finally the reversal path for annihilation of three skyrmions is calculated as shown in Fig. 6. With the standard parameter of the string method with $\omega$ = 0 the energy drastically increases in the vicinity of the saddle point. As a consequence the saddle point configuration cannot be properly resolved and the simulation does not converge. In order to overcome this problem the density of images is scaled with an energy norm using $\omega$ = 10. It can be seen that this approach is successful and the states around the saddle point can be resolved well. The saddle point configuration is a state where the two skyrmions in the bottom layer start to merge Fig. 6 (C). Since even in this annihilation process the topological charge changes the formation of a Bloch point occurs. Hence, the micromagnetic simulation is not used to provide the detailed energy barrier but it is used in order to provide a lower limit of the energy barrier since the chosen mesh size is significantly larger than the lattice constant. Since the obtained energy barrier of this annihilation processes is $\Delta E$ = 445 $k_B T_{300}$ and significantly larger than the annihilation via the boundary this estimate is sufficient to conclude that this annihilation process is not the most critical one.

To conclude a concept is proposed that shows how information can be reliably stored in two wires containing skyrmions. The bottom wire is used to generate dynamically moving pinning sites. These pinning sites are used to store the information by the skyrmion positions in the top layer at well defined distances. Due to the repulsive forces of the skyrmions in the bottom layer, all skyrmions in the shift register will move with constant distance between them. It is shown that the energy barrier for annihilation of skyrmions is larger than $\Delta E = 55\, k_B T_{300}$. This concept solves the significant problem of distributions of pinning times and currents that are a significant error source in skyrmion race track devices[12]. The areal density of the presented structure is 0.04 Tbit/inch², which is significantly smaller than the areal density of state-of-the-art magnetic recording devices [21] of 1.4 Tbit/inch². Hence, even if the material parameters are optimized in order to realize smaller thermally stable skyrmions it will be extremely challenging that skyrmion memories can compete with the areal density of existing magnetic storage devices such as hard disks.

The financial support by the Austrian Federal Ministry of Science, Research and Economy and the National Foundation for Research, Technology and Development as well as the Austrian Science Fund (FWF) under Grant Nos. F4112 SFB ViCoM, I2214-N20 and the Vienna Science and Technology Fund (WWTF) under Grant No. MA14-044 is acknowledged.

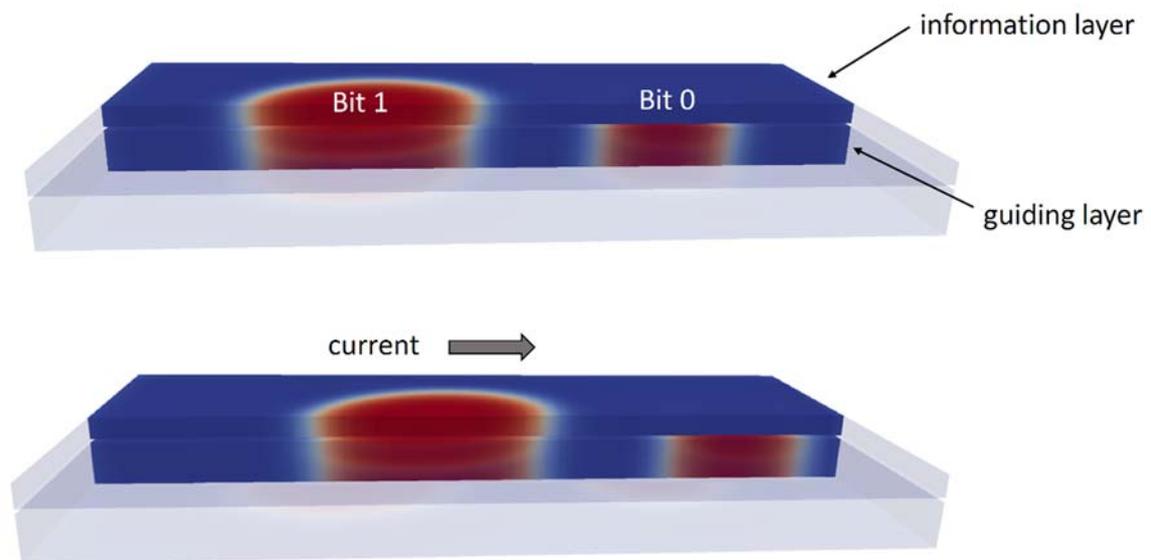

Fig. 1: Schematic view of the working principle of the gap protecting skyrmion device. The structure consists of two exchange decoupled layers. The top layer (information layer) stores the information by the presence and absence of a skyrmion. In order to keep the distance between these states, they are coupled via strayfield interaction to the bottom layer (guiding layer). This layer contains skyrmions at each bit position that are kept constant in distance due to their repulsive force. For writing and reading current pulses are applied until the bottom skyrmions move by one bit position.

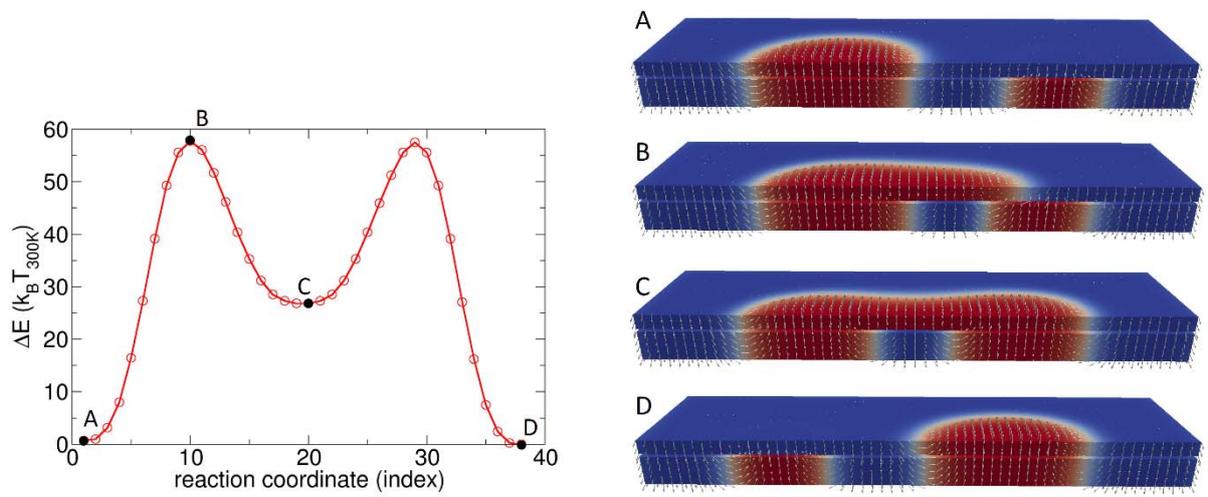

Fig. 2: Energy barrier of bit error process, where the stored information moves to the next dynamical pinning site. The thickness of the guiding layer is $t_{bottum}$ = 10 nm and the information layer $t_{top}$ = 5 nm. These two layers are separated by a 1 nm non-magnetic layer. The lateral dimensions are 180 nm x 90 nm.

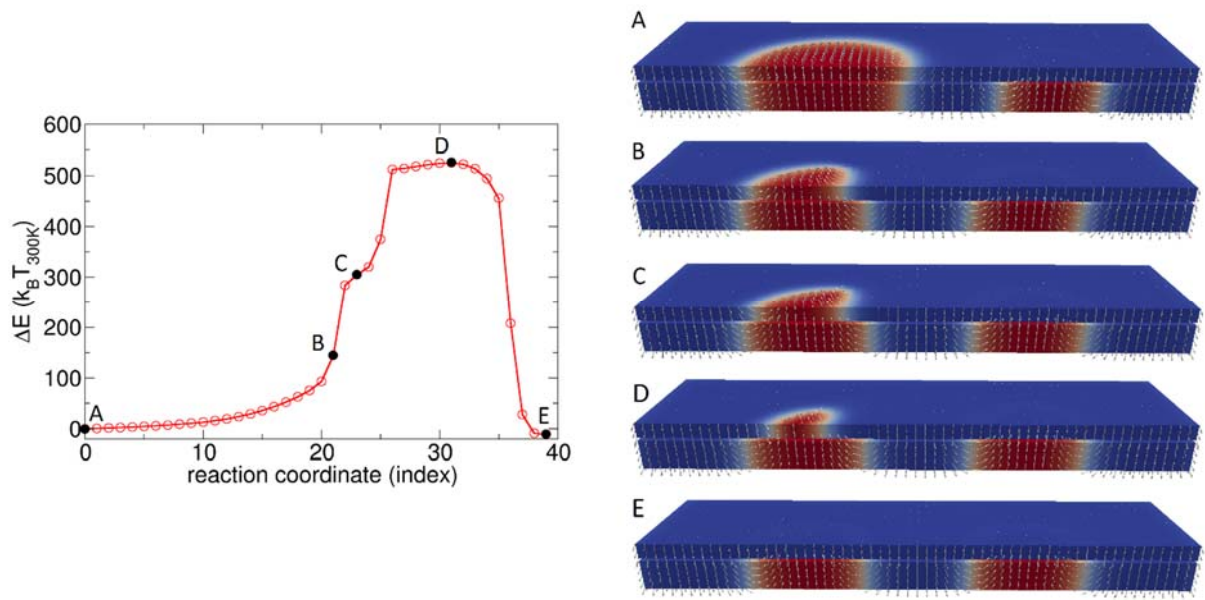

Fig. 3: *Energy barrier of bit error process, where the skyrmion in the information layer is annihilated via the formation of a Bloch point.*

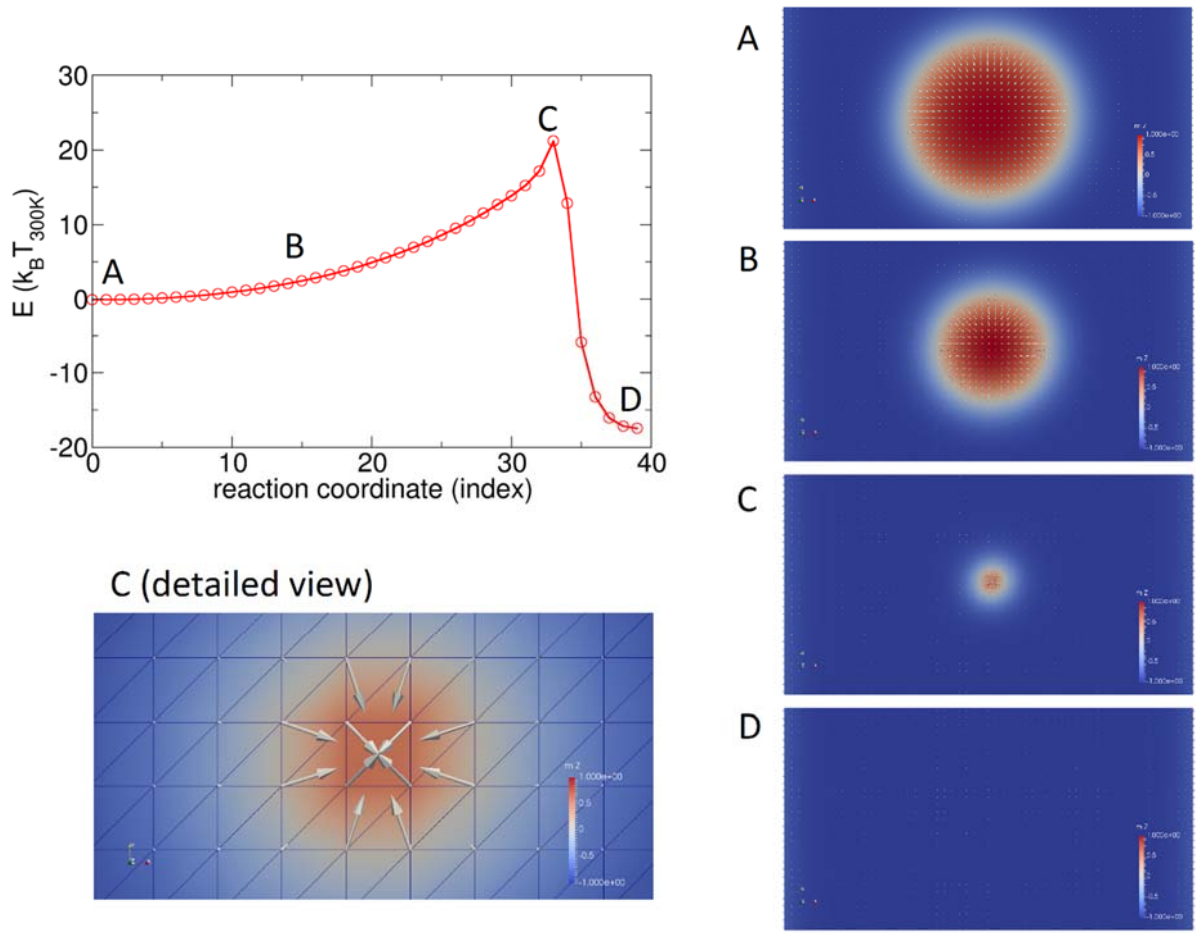

Fig. 4: Detailed calculation of the annihilation of a single skyrmion via the formation of a Bloch point. Only one magnetic layer with a thickness *t* = 0.6 nm is simulated. The lateral dimensions are 90 nm x 90 nm. The discretization mesh size is 1.6 nm.

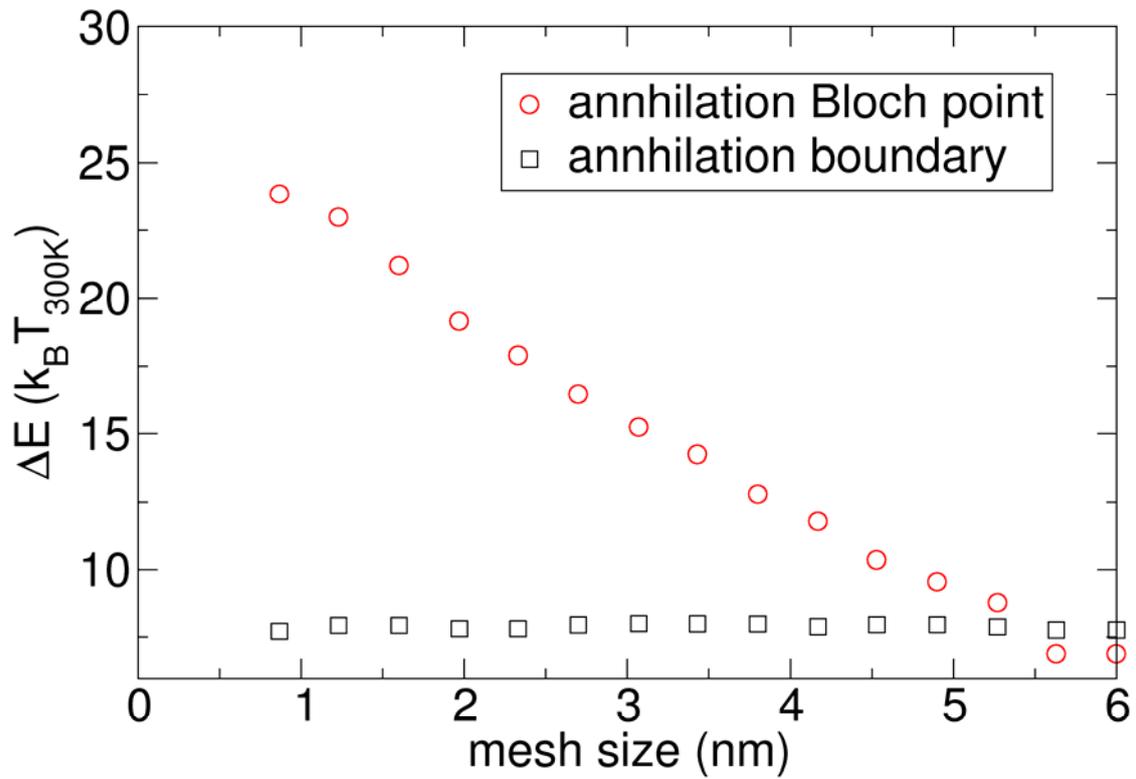

Fig. 5: Mesh size dependence of two annihilation processes of a skyrmion. (red circles) Annihilation within the structure occurs via formation of a Bloch point as shown in Fig. 4. The energy of the Bloch point depends on the discretization. (black squares) annihilation of the skyrmion via the boundary. The energy barrier does not depend on the mesh size, since no Bloch point is formed.

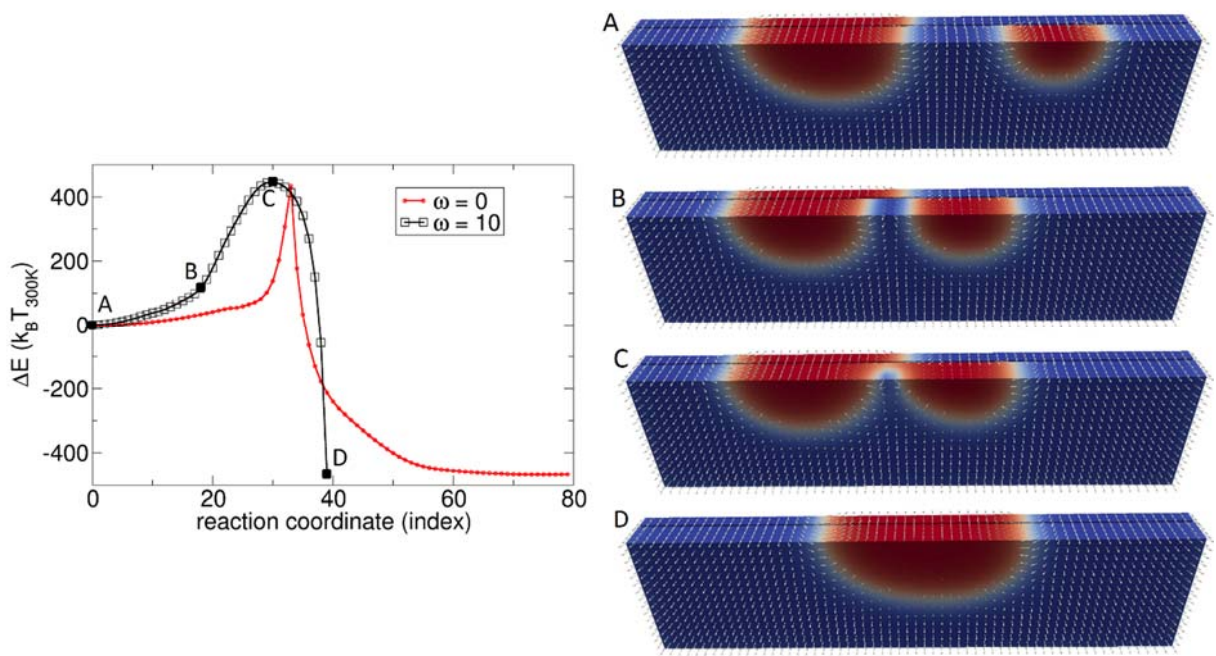

Fig. 6: *Energy barrier of bit error process, where three skyrmions are annihilated to two skyrmions. In order to increase the density of the images besides the standard parameter ω = 0 (red), a simulation with ω = 10 is performed (black).*